\documentstyle[12pt]{article}
\def\al{\alpha}
\def\be{\beta}
\def\ga{\gamma}
\def\de{\delta}
\def\ep{\epsilon}

\def\th{\theta}

\def\ka{\kappa}
\def\la{\lambda}

\def\si{\sigma}

\def\ph{\phi}

\def\ps{\psi}

\def\Ga{\Gamma}

\def\mn{{\mu\nu}}

\def\cl{{\cal L}}

\def\vev#1{\langle {#1}\rangle}

\def\fr#1#2{{{#1} \over {#2}}}
\def\frac#1#2{{\textstyle{{#1}\over {#2}}}}
\def\half{{\textstyle{1\over 2}}}
\def\lsim{\mathrel{\rlap{\lower4pt\hbox{\hskip1pt$\sim$}}
    \raise1pt\hbox{$<$}}}
\def\gsim{\mathrel{\rlap{\lower4pt\hbox{\hskip1pt$\sim$}}
    \raise1pt\hbox{$>$}}}
\def\sqr#1#2{{\vcenter{\vbox{\hrule height.#2pt
         \hbox{\vrule width.#2pt height#1pt \kern#1pt
         \vrule width.#2pt}
         \hrule height.#2pt}}}}

\def\lrprt{\stackrel{\leftrightarrow}{\partial}}

\def\lrDmu{\stackrel{\leftrightarrow}{D_\mu}}
\def\lrDnu{\stackrel{\leftrightarrow}{D^\nu}}

\def\Re{\hbox{Re}\,}

\newcommand{\beq}{\begin{equation}}
\newcommand{\eeq}{\end{equation}}
\newcommand{\bea}{\begin{eqnarray}}
\newcommand{\eea}{\end{eqnarray}}
\newcommand{\rf}[1]{(\ref{#1})}

\renewenvironment{thebibliography}[1]
 { \rm
   \begin{list}{\arabic{enumi}.}
    {\usecounter{enumi} \setlength{\parsep}{0pt}
     \setlength{\itemsep}{3pt} \settowidth{\labelwidth}{#1.}
     \sloppy
    }}{\end{list}}
 
\begin{document}
\titlepage
 
\vglue 1cm
	    
\begin{center}
{{\bf Possible Spontaneous Breaking of Lorentz and CPT Symmetry 
\\}
\vglue 1.0cm
{Don Colladay \\} 
\bigskip
{\it The College of Wooster,\\}
\medskip
{\it Wooster, OH, 44691, U.S.A.\\}
 
\vglue 0.8cm
}
\vglue 0.3cm
 
\end{center}
 
{\rightskip=3pc\leftskip=3pc\noindent
One possible ramification of unified theories of nature such as string 
theory that may underlie the 
conventional standard model is the possible 
spontaneous breakdown of Lorentz and CPT symmetry.
In this talk, the formalism for inclusion of such effects into a 
low-energy effective field theory is presented.
An extension of the standard model that includes Lorentz- and 
CPT-breaking terms is developed.
The restriction of the standard model extension to the QED sector is 
then discussed.
}

\vskip 1 cm

\begin{center}
{\it For the Proceedings of 
Non-Accelerator New Physics, June 28 -- July 3, 1999\\}
\end{center}

\baselineskip=20pt
 
\vglue 0.6cm
{\bf \noindent I. Introduction and Motivation}
\vglue 0.4cm

Virtually all modern particle physics theories are constructed using 
Lorentz invariance as a basic axiom.
Local, point-particle field theories, coupled with this assumed 
Lorentz invariance along with some mild technical assumptions 
leads one to conclude that CPT must also be preserved.\cite{cptthm}
The standard model, its supersymmetric extensions and grand unified 
models are all of this type.

However, if the fundamental theory underlying the standard model 
is constructed using nonlocal objects such as strings, 
Lorentz symmetry may be spontaneously broken in the low-energy 
limit of the full theory.
An explicit mechanism of this type 
has been proposed in the context of string theory.\cite{ks,kp1}
The Lorentz- and CPT-violating terms are generated when tensor fields 
gain vacuum expectation values through spontaneous symmetry breaking.

The approach adopted here is to use the mechanism of spontaneous symmetry 
breaking to generate a list of possible Lorentz violating interactions between 
standard model fields.
The standard model extension is then constructed by selecting 
those terms satisfying SU(3)$\times$SU(2)$\times$U(1) gauge invariance
and power-counting renormalizability.\cite{ck1}
By only using the property of spontaneous symmetry breaking and not 
referring to explicit details of the underlying theory, we are able 
to construct a general model of Lorentz breaking in the context
of the conventional standard model.

Many experimental tests of Lorentz and CPT invariance have been 
performed, 
so it is useful to have a general theory with explicit parameters 
that can be used to relate the various experiments as well as motivate 
new ones.
For example, 
high precision measurements involving atomic systems, 
\cite{ptrap,bkr}
clock comparisons, \cite{cc} and neutral meson oscillations 
\cite{kexpt,bexpt}
provide stringent 
tests of Lorentz and CPT symmetry.
The implications of CPT-violating terms on baryogenesis have also 
been investigated.\cite{bert}

To describe spontaneous Lorentz and CPT breaking, it is convenient 
to first review the Higgs mechanism in the standard model.
Conventional spontaneous symmetry breaking occurs in the 
Higgs sector of the standard model 
where the Higgs field obtains an expectation value, 
thereby partially 
breaking SU(2)$\times$U(1) gauge invariance.
This happens because an assumed potential for the Higgs field is 
minimized at some nonzero value of the field.

As an example, consider a simple Lagrangian describing a single
fermion field $\psi$ and a single scalar field $\phi$ of the form
\beq
{\cal{L}} = {\cal{L}}_{0} - {\cal{L}}^{\prime} 
\quad ,
\eeq
where
\beq
{\cal L}^{\prime} \supset {\la} \phi
\overline{\psi} \psi + \rm{h.c.}
- (\phi^{\dagger}\phi - a^{2})^{2}
\quad .
\label{higgsmech}
\eeq
A nonzero vacuum expectation value $\vev{\phi}$ for the 
scalar field will minimize the energy, 
hence generating a mass for the fermion of $m_{f}=\la \vev{\phi}$.
This expectation value of the scalar field breaks 
SU(2)$\times$U(1) gauge invariance because $\vev{\phi}$ no
longer transforms in a nontrivial way under this gauge group.
Lorentz symmetry is maintained in this case because $\vev{\phi}$
and $\phi$ are both scalars under the action of the Lorentz group. 

Suppose instead that a \it tensor \rm $T$ gains a nonzero vacuum 
expectation value,
$\vev{T}$.
Lorentz invariance is spontaneously broken in this case.
To see how this form of symmetry breaking might occur, consider a Lagrangian
describing a fermion $\psi$ and a tensor $T$ of the form
\beq
{\cal{L}} = {\cal{L}}_{0} - {\cal{L}}^{\prime} 
\quad ,
\eeq
where
\beq
{\cal L}^{\prime} \supset \fr {\la}{M^k} {T} \cdot \overline{\psi} 
\Ga (i \partial )^k \psi + \rm{h.c.} + V(T)
\quad .
\eeq
In this expression, $\la$ is a dimensionless coupling, 
$M$ is a heavy mass scale of the underlying theory,
$\Ga$ denotes a general gamma matrix structure in the Dirac algebra,
and $V(T)$ is a potential for the tensor field. (indices are
suppressed for notational simplicity)
Terms contributing to $V(T)$ are precluded from
conventional renormalizable four-dimensional field theories,
but may arise in the low-energy limit 
of a more general theory such as string theory.
\cite{ks}

If the potential $V(T)$ is such that it has a nontrivial minimum, a 
vacuum expectation value $\vev{T}$ will be generated for the tensor 
field.
There will then be a term of the form
\beq
{\cal L}^{\prime} \supset \fr {\la}{M^k} \vev{T} \cdot \overline{\psi} 
\Ga (i \partial )^k \chi + \rm{h.c.} 
\quad ,
\label{vevt}
\eeq
present in the Lagrangian after spontaneous symmetry breaking occurs.
These terms can break Lorentz invariance and various 
discrete symmetries C, P, T, CP, and CPT.

\vglue 0.6cm
{\bf \noindent II. Relativistic Quantum Mechanics and Field Theory}
\vglue 0.4cm

To develop theoretical techniques for treating generic terms of the 
type given in Eq.~\rf{vevt}, we first study a specific example.
The example presented here involves a lagrangian for a single fermion 
field containing Lorentz-violating terms with no derivative couplings 
($k=0$) that also violate CPT.

We proceed by listing the possible gamma-matrix structures that could 
arise within such a term:
\beq
\Ga \sim \{1, \ga^{\mu}, \ga^5 \ga^{\mu}, 
\sigma^{\mn}, \ga^5\}
\quad .
\label{gambasis}
\eeq
The condition that a fermion bilinear with no derivative couplings 
violates CPT is equivalent to the requirement that $\Ga$ be chosen 
such that $\{\Ga,\ga^{5}\}=0$.
Half the matrices in Eq.~\rf{gambasis} 
satisfy this condition: $\Ga \sim \ga^{\mu}$ and 
$\Ga \sim \ga^{5}\ga^{\mu}$.
The contribution to the lagrangian from these terms can be written as
\beq
{\cal L}_{a}^{\prime} \equiv a_{\mu} \overline{\psi} \ga^{\mu} \psi \quad , 
\quad
{\cal L}_{b}^{\prime} \equiv b_{\mu} \overline{\psi} \ga_5 \ga^{\mu} \psi 
\quad ,
\label{abmu}
\eeq
where $a_{\mu}$ and $b_{\mu}$ are constant coupling coefficients that 
parameterize the tensor expectation values and relevant coupling 
constants arising in Eq.~\rf{vevt}.
These parameters are assumed suppressed with respect to other 
physically relevant energy scales in the low-energy effective theory 
in order to be in agreement with current experimental bounds.

Including these contributions from the spontaneous symmetry breaking 
mechanism into a theory containing a free Dirac fermion yields a model 
lagrangian of 
\beq
{\cal L} = \fr i 2 \overline{\ps} \ga^{\mu} \lrprt_\mu \psi 
- a_{\mu} \overline{\psi} \ga^{\mu} \psi 
- b_{\mu} \overline{\psi} \ga_5 \ga^{\mu} \psi 
- m \overline{\psi} \psi
\quad .
\label{modlag}
\eeq
Several features of this modified theory are immediately apparent 
upon inspection.
The first feature is that the lagrangian is hermitian, 
thereby leading to a theory obeying conventional quantum 
mechanics, conservation of probability and unitarity.
The second feature is that translational invariance implies 
the existence of a conserved energy and momentum.
This conserved four-momentum is explicitly constructed as
\beq
P_\mu = \int d^3 x \Theta^0_{\ \mu} = 
\int d^3 x \frac 1 2 i \overline{\psi} \ga^0 \lrprt_\mu \psi
\quad ,
\label{emom}
\eeq
just as in the conventional case.
The third feature is that the Dirac equation resulting from 
Eq.~\rf{modlag} is linear in the 
fermion field allowing an exact solution to the free theory.
Finally, a global U(1) invariance of the model lagrangian implies 
the existence of a conserved current 
$j_{\mu}=\overline{\psi}\ga^{\mu} \psi$.

The Dirac equation obtained by variation of Eq.~\rf{modlag} 
with respect to the fermion field is
\beq
(i \ga^{\mu} \partial_{\mu} - a_\mu \ga^\mu - 
b_{\mu} \ga_5 \ga^\mu - m) \psi = 0
\quad .
\eeq
Due to the linearity of the equation, plane-wave solutions 
\beq
\psi(x) = e^{\pm i p_{\mu} x^{\mu}} w(\vec{p}) 
\quad ,
\eeq
are used to solve the equation exactly.
Substitution of the plane-wave solution 
into the modified Dirac equation yields
\begin{eqnarray}
(\pm p_{\mu} \ga^{\mu} - a_{\mu} \ga^{\mu} - 
b_{\mu} \ga_5 \ga^{\mu} - m) w(\vec{p}) &\equiv& M_{\pm} w(\vec{p}) 
\nonumber \\ &=& 0 
\quad .
\end{eqnarray}
A nontrivial solution exists only if $Det M_{\pm} = 0$.
This imposes a condition on $p^0(\vec{p}) \equiv E(\vec{p})$, 
hence generating a dispersion relation for the fermion.

The general solution involves finding the roots of a 
fourth-order polynomial equation.
The solutions can be found algorithmically, but the resulting
solution is complex and not very illuminating.
For simplicity we consider only the special case of $\vec{b}=0$ here.
The exact dispersion relations for this case are
\beq
E_{+}(\vec{p}) = \left[ m^2 + (|\vec{p} - \vec{a}| \pm b_0)^2 \right]^{1/2} 
+ a_0 \quad ,
\eeq
\beq
E_{-}(\vec{p}) = \left[ m^2 + (|\vec{p} + \vec{a}| \mp b_0)^2 \right]^{1/2} 
- a_0 \quad .
\eeq
Examination of the above energies reveals several qualitative effects of the 
CPT-violating terms.
The usual four-fold energy degeneracy of spin-$\frac 1 2$ 
particles and antiparticles 
is removed by the $a_{\mu}$ and $b_{0}$ terms.
The particle-antiparticle energy degeneracy is broken by $a_{\mu}$ and the 
helicity degeneracy is split by $b_{0}$.
The corresponding spinor solutions $w(\vec p)$ have been explicitly calculated, 
forming an orthogonal basis of states as expected.

An interesting feature of these solutions is the unconventional 
relationship that exists between momentum and velocity.
A wave packet of positive helicity particles with four momentum 
$p^{\mu}=(E,\vec{p})$ has an expectation value of the velocity operator 
$\vec{v} = i[H,\vec{x}] = \ga^0 \vec{\ga}$ of
\beq
\vev{\vec{v}} = 
\vev{ \fr {(|\vec{p} - \vec{a}| - b^0)} {(E - a^0)}
         \fr {(\vec{p} - \vec{a})} {|\vec{p} - \vec{a}|} }
         \quad .
\eeq
Examination of the above velocity using a general dispersion relation
reveals that $|v_{j}| < 1$ for arbitrary $b_{\mu}$,
and that the limiting velocity as $\vec{p} \rightarrow \infty$ is 1.
This implies that the effects of the CPT violating terms are mild 
enough to preserve causality in the theory.
This will be verified independently using the perspective of field 
theory that will now be developed.

To quantize the theory, the general expansion for $\psi$ in terms of 
its spinor components given by
\begin{eqnarray}
\ps (x) & = & \int \fr {d^3 p} {(2 \pi )^3} \sum_{\al = 1}^{2} \left[
\fr m {E_u^{(\al)}} b_{(\al)} (\vec{p})
e^{-i p_u^{(\al)} \cdot x} u^{(\al)} (\vec{p}) \right. \nonumber \\ 
&& \left. \qquad \qquad \qquad \qquad
+ \fr m {E_v^{(\al)}} d^*_{(\al)} (\vec{p}) 
e^{i p_v^{(\al)} \cdot x} v^{(\al)} (\vec{p}) \right] 
\quad ,
\end{eqnarray}
is promoted to an operator acting on a Hilbert space 
of basis states.
The energy is calculated from Eq.~\rf{emom} using conventional 
normal ordering.
The result is a positive definite quantity (for $|a^{0}| <m$)
provided the following nonvanishing anticommutation relations are imposed
on the creation and annihilation operators:
\begin{eqnarray}
\{b_{(\al)} (\vec{p}), b^{\dagger}_{(\al^{\prime})} (\vec{p}^{~\prime}) \} 
& = & (2 \pi)^3
\fr {E_u^{(\al)}} {m}
\de_{\al \al^{\prime}}
\de^3 (\vec{p} - \vec{p}^{~\prime})
\quad ,
\nonumber \\
\{d_{(\al)} (\vec{p}), d^{\dagger}_{(\al^{\prime})} (\vec{p}^{~\prime}) \} 
& = & (2 \pi)^3
\fr {E_v^{(\al)}} {m}
\de_{\al \al^{\prime}}
\de^3 (\vec{p} - \vec{p}^{~\prime})
\quad .
\end{eqnarray}
The resulting equal-time anticommutators for the fields are
\begin{eqnarray}
\{ \psi_{\al}(t,\vec{x}), \psi_{\be}^{\dagger}(t,\vec{x}^{\prime})\} 
& = & \de_{\al \be} \de^{3} (\vec{x} - \vec{x}^{\prime})
\quad , 
\nonumber \\ 
\{ \psi_{\al}(t,\vec{x}), \psi_{\be}(t,\vec{x}^{\prime})\} & = & 0 
\quad ,
\nonumber \\
\{ \psi_{\al}^{\dagger}(t,\vec{x}), \psi_{\be}^{\dagger}(t,\vec{x}^{\prime})\} 
& = & 0
\quad .
\end{eqnarray}
These relations show that conventional Fermi statistics remain 
unaltered in the presence of Lorentz- and CPT-violating terms.

The conserved charge $Q$ and four-momentum $P^{\mu}$ are computed as
\begin{eqnarray}
Q & = & \int \fr {d^3 p} {(2 \pi)^3}
\sum_{\al = 1}^2 \left[
\fr m {E_u^{(\al)}}
b^{\dagger}_{(\al)} (\vec{p}) b_{(\al)} (\vec{p}) - \fr m {E_v^{(\al)}}
d^{\dagger}_{(\al)} (\vec{p}) d_{(\al)} (\vec{p}) \right] \quad ,\\
P_{\mu} & = & \int \fr {d^3 p} {(2 \pi)^3} \sum_{\al = 1}^2 \left[
\fr m {E_u^{(\al)}} p^{(\al)}_{u \mu}
b^{\dagger}_{(\al)} (\vec{p}) b_{(\al)} (\vec{p}) \right. 
\nonumber \\ 
& & \qquad \qquad \qquad \qquad \qquad \left.
+ \fr m {E_v^{(\al)}} p^{(\al)}_{v \mu}
d^{\dagger}_{(\al)} (\vec{p}) d_{(\al)} (\vec{p}) \right]
\quad .
\end{eqnarray}
From these expressions we see that the charge of the fermion is 
unperturbed and the energy and momentum satisfy the same  
relations that are found using relativistic quantum mechanics.

Causality is governed by the anticommutation relations 
of the fermion fields at unequal times.  
Explicit integration in the special case $\vec{b}=0$ proves that
\beq
\{\psi_{\al}(x), \overline{\psi}_{\be}(x^{\prime})\} = 0
\quad ,
\eeq
for spacelike separations $(x - x^{\prime})^2 < 0$.
The above result shows that physical observables separated by spacelike 
intervals will in fact commute (for case $\vec b = 0$).
This agrees with our previous results obtained by examination of the 
velocity using relativistic quantum mechanics.

Next, the problem of extending the free field theory to interacting 
theory is addressed.
Much of the conventional formalism developed for perturbative 
calculations in the interacting theory carries over directly to the 
present case.
The main reason that these techniques work is that the Lorentz 
violating modifications which are introduced are linear in the fermion fields.
The main result is that the usual Feynman rules apply provided  
the Feynman propagator is modified to
\beq
S_F(p) = \fr i {p_{\mu} \ga^{\mu} - a_{\mu} \ga^{\mu} - 
b_{\mu} \ga_5 \ga^{\mu} - m}
\quad ,
\eeq
and the exact spinor solutions of the modified free fermion theory 
are used on the external legs of the diagrams.

\vglue 0.6cm
{\bf \noindent III. Extension of The Standard Model}
\vglue 0.4cm

In this section the question of how to apply spontaneous symmetry
breaking to generate Lorentz-violating terms using standard model
fields is addressed.
Our approach involves consideration of all possible terms that can arise from
spontaneous symmetry breaking that satisfy power-counting 
renormalizability and preserve the SU(3)$\times$SU(2)$\times$U(1) 
gauge invariance of the standard model.\cite{ck1}
Even with these constraints, 
terms are found to contribute to all sectors of the standard 
model.
In listing the terms here, the Lorentz violating terms are 
classified according to their properties under the CPT transformation.

In the lepton sector the 
left- and right-handed multiplets are defined as
\beq
L_A = \left( \begin{array}{c} \nu_A \\ l_A \end{array} \right)_L
\quad , \quad
R_A = (l_A)_R
\quad ,
\eeq
where $A = 1, 2, 3$ labels the flavor:
\beq
l_A \equiv (e, \mu, \tau) \quad , \quad
\nu_A \equiv (\nu_e, \nu_{\mu}, \nu_{\tau}) 
\quad .
\eeq

The Lorentz-violating terms that satisfy the required properties are
\begin{eqnarray}
\cl^{\rm CPT-even}_{\rm lepton} &=& 
\half i (c_L)_{\mu\nu AB} \overline{L}_A \ga^{\mu} \lrDnu L_B
\nonumber\\ &&
+ \half i (c_R)_{\mu\nu AB} \overline{R}_A \ga^{\mu} \lrDnu R_B
\quad , 
\\ && \nonumber \\
\cl^{\rm CPT-odd}_{\rm lepton} & = &
- ({a}_{L})_{\mu A B} ~ \overline{L}_A \ga^{\mu} L_B \nonumber \\ 
& & - ({a}_{R})_{\mu A B} ~ \overline{R}_A \ga^{\mu} R_B 
\quad .
\label{leptonlv}
\end{eqnarray}
In the above expression
$c_{\mu\nu}$ and $a_{\mu}$ are constant coupling coefficients related 
to the background expectation values of the relevant tensor 
fields, and $D^{\mu}$ is the conventional covariant derivative.

The final form of the standard model terms is different because the 
SU(2)$\times$U(1) symmetry is broken by the Higgs mechanism.  
Once this breaking occurs, the fields in Eq.~\rf{leptonlv}
are rewritten in terms of the physical Dirac spinors corresponding 
to the observed leptons and neutrinos.
As an example, the CPT-odd lepton terms become
\begin{eqnarray}
\cl^{\rm CPT-odd}_{\rm lepton} & = &
-(a_\nu)_{\mu A B} ~ \overline{\nu}_{A}
\frac 1 2 (1 + \ga_5)\ga^{\mu} \nu_B \nonumber \\
& & - (a_l)_{\mu A B} ~ \overline{l}_A \ga^{\mu} l_B \nonumber \\ 
& & - (b_l)_{\mu A B} ~ \overline{l}_A \ga_5 \ga^{\mu} l_B 
\quad .
\label{lepcpt}
\end{eqnarray}
Note that $b_{\mu}$ coupling coefficients arise in 
the process of combining the right- and left-handed fields into Dirac 
spinors.

If we now examine the first generation electron contribution 
corresponding to $A=B=1$, we find the terms
\beq
\cl^{\rm CPT-odd}_{\rm lepton} 
\supset - (a_l)_{\mu 1 1}~\overline{e} \ga^{\mu} e 
- (b_l)_{\mu 1 1} ~\overline{e}\ga_5 \ga^{\mu} e 
\quad .
\eeq
These terms are exactly the form as the contributions to the 
model lagrangian of Eq.~\rf{modlag}
that were analyzed in the previous section.
The relativistic quantum mechanics and field theoretic techniques that 
were developed to handle these terms are therefore directly applicable to 
electrons.
Terms in Eq.~\rf{lepcpt} of the form $A \ne B$ contribute small lepton 
flavor-changing amplitudes.

The construction of the standard model extension in the quark sector is 
similar to that in the lepton sector. 
The main difference is that corresponding right-handed quark fields 
are present for each left-handed field unlike the case in 
the lepton sector.
The left- and right-handed quark multiplets are denoted
\beq
Q_A = \left( \begin{array}{c} u_A \\ d_A \end{array} \right)_L
\quad , \quad
\begin{array}{c}
U_A = (u_A)_R \\
D_A = (d_A)_R
\end{array}
\quad ,
\eeq
where $A = 1, 2, 3$ labels quark flavor
\beq
u_A \equiv (u,c,t) \quad , \quad
d_A \equiv (d,s,b) \quad .
\eeq
The Lorentz-violating terms in the quark sector are of the same form 
as in the lepton sector.
The diagonal $A=B$ terms are again of the same form as Eq.~\rf{abmu}.
The quark $a_{\mu}$ terms are particularly interesting because they can lead 
to observable CPT-violating effects in neutral meson systems.\cite{kost}

In the Higgs sector, there are contributions involving two Higgs 
fields, and generalized Yukawa coupling terms involving a single Higgs 
and two fermion fields.
The Lorentz-violating terms that are quadratic in the Higgs fields are
\begin{eqnarray}
\cl^{\rm CPT-even}_{\rm Higgs} &=&
\half (k_{\ph\ph})^{\mu\nu} (D_\mu\ph)^\dagger D_\nu\ph 
+ {\rm h.c.}
\nonumber\\ &&
-\half (k_{\ph B})^{\mu\nu} \ph^\dagger \ph B_{\mu\nu}
\nonumber\\ &&
-\half (k_{\ph W})^{\mu\nu} \ph^\dagger W_{\mu\nu} \ph 
\quad ,
\end{eqnarray}
\beq
\cl^{\rm CPT-odd}_{\rm Higgs}
= i (k_\ph)^{\mu} \ph^{\dagger} D_{\mu} \ph + {\rm h.c.} 
\quad ,
\eeq
where $W_{\mu\nu}$ and $B_{\mu\nu}$ are the field strengths for the 
SU(2) and U(1) gauge fields and the various $k$ parameters are 
coupling constants related to tensor expectation values.

The Yukawa type terms involving one Higgs field are
\begin{eqnarray}
\cl^{\rm CPT-even}_{\rm Yukawa} = 
&& - \half 
\left[
(H_L)_{\mu\nu AB} \overline{L}_A \ph \si^{\mu\nu} R_B
\right.
\nonumber\\ &&
\left.
+(H_U)_{\mu\nu AB} \overline{Q}_A \ph^c \si^{\mu\nu} U_B 
\right.
\nonumber\\ &&
\left.
+(H_D)_{\mu\nu AB} \overline{Q}_A \ph \si^{\mu\nu} D_B
\right]
+ {\rm h.c.}
\quad ,
\end{eqnarray}
where the $H$ parameters are related to tensor expectation values.

One interesting result of including these terms into the standard 
model is a modification of the conventional SU(2)$\times$U(1) 
breaking.
When the full static potential is minimized, the $Z^{0}$ boson gains an 
expectation value of
\beq 
\vev{Z_\mu^0} = \fr 1 q {\sin2\th_W} 
(\Re \hat k_{\ph\ph})^{-1}_{\mu\nu} k^\nu_\ph
\quad ,
\eeq
where $\hat k_{\ph\ph}^{\mu\nu} = \eta^{\mu\nu} + 
k_{\ph\ph}^{\mu\nu}$, $q$ is the electric charge,
and $\th_{W}$ is the weak mixing angle.
If the CPT-odd term $k_{\ph}$ vanishes then 
$\vev{Z_\mu^0}=0$.  This is reasonable since a nonzero value of 
$\vev{Z_\mu^0}$ violates CPT symmetry.

The gauge sector is the final sector to be examined.  
The various Lorentz-breaking terms satisfying the relevant criteria are
\begin{eqnarray}
\cl^{\rm CPT-even}_{\rm gauge} &=&
-\half (k_G)_{\ka\la\mu\nu} {\rm Tr} (G^{\ka\la}G^{\mu\nu})
\nonumber\\ &&
-\half (k_W)_{\ka\la\mu\nu} {\rm Tr} (W^{\ka\la}W^{\mu\nu})
\nonumber\\ &&
-\frac 1 4 (k_B)_{\ka\la\mu\nu} B^{\ka\la}B^{\mu\nu}
\quad ,
\label{gaugee}
\end{eqnarray}
\begin{eqnarray}
\cl^{\rm CPT-odd}_{\rm gauge} & = &
k_{3\ka} \ep^{\ka\la\mu\nu}
{\rm Tr} (G_\la G_{\mu\nu} + \frac {2i} 3 G_\la G_\mu G_\nu) 
\nonumber \\ 
& +& k_{2\ka} \ep^{\ka\la\mu\nu}
{\rm Tr} (W_\la W_{\mu\nu} + \frac {2i} 3 W_\la W_\mu W_\nu) 
\nonumber \\ 
& +& k_{1\ka} \ep^{\ka\la\mu\nu} B_\la B_{\mu\nu} 
\quad .
\label{gaugeo}
\end{eqnarray}
In these expressions, the $k$ terms are constant coupling constants
and the $G^{\mu\nu}$, $W^{\mu\nu}$, and $B^{\mu\nu}$ are the field 
strengths for the SU(3), SU(2), and U(1) gauge fields respectively.

The CPT-odd terms can generate negative contributions to the conserved 
energy \cite{cfj}, hence creating an instability in the theory.  
It is therefore desirable to set these coefficients to zero, 
provided they remain zero at the quantum level.  
This procedure has been carried out to the 
one-loop level by utilizing an anomaly cancellation mechanism that 
must be inherited from any consistent theory underlying the standard 
model.\cite{ck1}
This point is discussed further in the following section.

\vglue 0.6cm
{\bf \noindent IV. QED Restriction}
\vglue 0.4cm

We now restrict our attention to the theory of electrons and photons 
that results from the above extension of the standard model.
The conventional QED Lagrangian is 
\beq
\cl^{\rm QED}_{\rm electron} = 
\half i \overline{\ps} \ga^\mu \lrDmu \ps 
- m_e \overline{\ps} \ps
- \frac 1 4 F_{\mu\nu}F^{\mu\nu}
\quad ,
\eeq
where $\psi$ is the electron field, $m_{e}$ is its mass,
and $F^{\mu\nu}$ is the photon field strength tensor.

The CPT-even electron terms that violate Lorentz symmetry
in the full standard model extension are
\begin{eqnarray}
\cl^{\rm CPT-even}_{\rm electron} &=& 
- \half H_{\mu\nu} \overline{\ps} \si^{\mu\nu} \ps 
\nonumber\\ &&
+ \half i c_{\mu\nu} \overline{\ps} \ga^{\mu} \lrDnu \ps 
\nonumber\\ &&
+ \half i d_{\mu\nu} \overline{\ps} \ga_5 \ga^\mu \lrDnu \ps
\quad ,
\end{eqnarray}
where $H$, $c$, and $d$ are constant coupling coefficients.
The CPT-odd electron terms are
\beq
\cl^{\rm CPT-odd}_{\rm electron} = 
- a_{\mu } \overline{\ps} \ga^{\mu} \ps 
- b_{\mu} \overline{\ps} \ga_5 \ga^{\mu}\ps 
\quad ,
\eeq
where $a$ and $b$ are parameters analogous to those in Eq.~\rf{abmu} 
applied to electrons.

Experiments involving conventional QED tests can be used to place 
stringent bounds on the above violation parameters.
For example, Penning traps may be used to compare energy levels of 
$e^{-}$ and $e^{+}$ or $p$ and $\overline p$ orbits to constrain 
various combinations of parameters to few parta in $10^{20}$.\cite{bkr}
In addition, tests involving comparison of hydrogen and antihydrogen 
$1S-2S$ and hyperfine transitions can place comparable 
bounds on other combinations of parameters.\cite{bkr2}

The corrections to the photon from the gauge sector are given by
\beq
\cl^{\rm CPT-even}_{\rm photon} =
-\frac 1 4 (k_F)_{\ka\la\mu\nu} F^{\ka\la}F^{\mu\nu}
\quad ,
\eeq
and
\beq
\cl^{\rm CPT-odd}_{\rm photon} =
+ \half (k_{AF})^\ka \ep_{\ka\la\mu\nu} A^\la F^{\mu\nu}
\quad ,
\eeq
where the parameters $k_{F}$ and $k_{AF}$ are the appropriate linear 
combinations of parameters in Eqs.\rf{gaugee} and \rf{gaugeo} that 
result when the photon is defined as the unbroken U(1) electric force 
mediator.

A stringent limit of $(k_{AF})^{\mu} < 10^{-42}{\rm GeV}$ has been 
placed on the CPT-odd term using cosmological birefringence 
tests.\cite{cfj} 
Coupled with the theoretical difficulties involving negative 
contributions to the energy, this experimental bound 
indicates that this 
coefficient should be set identically to zero in the theory.
At first sight, radiative corrections appear to induce a nonzero term at the 
quantum level.
However, such corrections must cancel provided the underlying theory 
is anomaly free.

The only QED correction term with matching C, P, and T symmetry properties 
that contributes to $(k_{AF})^{\mu}$ is $b^{\mu}$.
The one-loop diagram produces an ambiguous, finite, and regularization 
dependent correction of $(k_{AF})^{\mu} = \zeta b^{\mu}$, where 
$\zeta$ is an arbitrary constant.\cite{ck1}
When this correction is summed over all fermion species, the 
contributions must cancel provided there is no amomaly in the full 
underlying theory.
A zero result to lowest order in $b^{\mu}$ has also been argued as a 
consistent choice using  
arguments based on the gauge invariance of the lagrangian.\cite{cg}
Several other recent works have shown similar results in various 
regularization schemes.\cite{rik1}

More recently, a calculation to all orders in $b^{\mu}$ using the 
exact modified propagator has been carried out.\cite{rik2}
Remarkably, the full result is the same as the correction generated 
by the linear term.
This means that the anomaly cancellation mechanism applies to all 
orders in $b^{\mu}$ and the coefficient $(k_{AF})^{\mu}$ remains 
zero at the quantum level.

The CPT-even terms are more interesting for several reasons.
First, the total canonical energy is positive provided the couplings 
are reasonably suppressed.  
Secondly, the contribution to cosmological birefringence is suppressed 
relative to the CPT-odd term.
Constraints of a few parts in $10^{23}$ have been obtained on the rotationally 
invariant term using cosmic-ray tests.\cite{cg} 
More general terms can be bounded to $k_{F}\le 10^{-28}$ 
using cosmological birefringence measurements.\cite{ck1}

\vglue 0.6cm
{\bf \noindent V. Summary}
\vglue 0.4cm

A framework has been 
presented that incorporates Lorentz- and CPT-violating effects into 
the context of conventional quantum field theory.
Using a generic spontaneous symmetry breaking mechanism as the source 
for these terms, an extension of the standard model that includes
Lorentz and CPT breaking was developed.
This extension preserves power-counting renormalizability and 
SU(3)$\times$SU(2)$\times$U(1) 
gauge invariance.
The parameters that have been introduced can be used to establish 
quantitative bounds 
on CPT- and Lorentz-breaking effects in nature.
Implications for electron and photon propagation in the QED sector 
were discussed.

\vglue 0.6cm
{\bf \noindent Acknowledgments}
This work was supported in part by the United States Department of 
Energy under grant number DE-FG02-91ER40661.

\vglue 0.6cm
{\bf \noindent REFERENCES}
\vglue 0.4cm

\end{document}